\documentclass[preprint,showpacs,preprintnumbers,amsmath,amssymb]{revtex4}

\def\p{\partial}

\def\ds{\displaystyle}

\def\H{{}^\star\!H}

\def\hd{\hat{\delta}}

\def\hx{\hat{\xi}}

\def\ch{{\cal{H}}}

\begin{document}

\title{M5-brane as a Nambu-Poisson geometry of a Multi D1-brane theory}

\author{A. De Castro}
\email{adecastr@pion.ivic.ve}
 \affiliation{Centro de Fi\'{\i}sica\\Instituto Venezolano de Investigaciones Cient\'{\i}ficas,
 Caracas}
\author{M. P. Garcia del Moral}
 \email{mgarcia@fis.usb.ve}
 \affiliation {Departamento de F\'{\i}sica,\\
Universidad Sim\'on Bol\'{\i}var,Caracas, e Instituto de F\'{\i}sica Te\'orica, Universidad
Aut\'onoma de Madrid, Madrid}
\author{I. Martin }
 \email{isbeliam@usb.ve}
\author{A. Restuccia}
\email{arestu@usb.ve} \affiliation{Departamento de F\'{\i}sica,
Universidad Sim\'on Bol\'{\i}var,\\  Caracas}

\date{\today}% It is always \today, today,
             %  but any date may be explicitly specified

\begin{abstract}
We introduce a Nambu-Poisson bracket in the geometrical description 
of the $D=11$ M5-brane. This procedure allows us, 
under some assumptions, to eliminate the local degrees of freedom of 
the antisymmetric field in the M5-brane Hamiltonian and to express it as 
a $D=11$ p-brane theory invariant under symplectomorphisms. A regularization of the M5-brane  
in terms of a multi D1-brane theory invariant under the $SU(N)\times SU(N)$ group in 
the limit when $N\rightarrow\infty$ is proposed. Also, a regularization for the  $D=10$ 
D4-brane in terms of a multi D0-brane theory is suggested. 
\end{abstract}

% PACS, the Physics and Astronomy
                             % Classification Scheme.
%\keywords{Suggested keywords}%Use showkeys class option if keyword
                              %display desired
\maketitle
%%%%%%%%%%%%%%%%%%%%%%%%%%%%%%%%%%%%%%%%%%%%%%%%%%%%%%%%%%%%%%%%%%%%%%%%%%%%%%%%%%%%

\section{Introduction}
Although the geometrical and algebraic structure of the  M5-brane
has been considered by several authors, it  is far from been
completely understood. There are natural questions concerning the
structure of the spectrum of its Hamiltonian that remain without
answer. In that sense an important aspect to analyze is the role of
the volume preserving diffeomorphisms as a residual gauge symmetry
of the canonical action once the light cone gauge has been fixed.
In the case of the $D=11$ Supermembrane, the residual gauge
symmetry is the area preserving diffeomorphisms group, which
coincides with the symplectomorpisms preserving a canonical
two-form $\omega$. In two dimensions $\omega$ may be expressed in
terms of the totally antisymmetric tensor density and the volume
element which is naturally introduced by the light cone gauge
fixing procedure. In this case, then, both groups coincide,
however, for higher dimensional p-branes or D-Branes, the
symplectomorphisms are only a subgroup of the volume preserving
diffeomorphisms. The understanding of this residual gauge symmetry
is directly related to the problem of finding an interesting
regularization of the M5-brane in terms of multi D0 or D1-branes.
That regularization was crucial in  understanding  the spectrum of
the $D=11$ Supermembrane. In that case, the regularization was
performed in terms of a quantum mechanical system which may be
interpreted as a $SU(N)$, $N\rightarrow\infty$, Yang-Mills theory
on zero spatial dimensions \cite{deWit:1989ct}or as a multi
D0-brane theory \cite{Banks:1997vh}. It was then understood that
the existence of a continuous spectrum from $0$ to $\infty$, in
the supersymmetric case, is directly related to the existence of
string-like spikes which may be attached to any membrane without
changing its energy. The supersymmetry was relevant since it
annihilates an effective basin shaped bosonic potential arising
from the zero point energy of the harmonic oscillators associated
to the membrane potential. It is already known
\cite{DeCastro:2001gp},\cite{DeCastro:2002vd} the existence of
4,3,2 and 1-branes spikes, which may be attached to the M5-brane
without changing its classical energy. This is also a general property  of
all p-Branes \cite{Nicolai:1998ic}. It is not known however its
relation to the spectrum of the M5-brane quantum Hamiltonian. Even more,
it is not known if there exists a suitable regularization for the
M5-brane. In this letter we start to analyze some of these questions.

We will show how the volume preserving generators and the M5-brane
Hamiltonian fits into a description where the world volume is
endowed with a Nambu-Poisson structure. This algebraic structure
was already proposed for p-branes in \cite{Hoppe:1997xp}. The
Nambu-Poisson structure over the world volume is directly related
to the canonical Poisson structure over the fields describing the
M5-brane. We will then consider the reduction of the volume
preserving diffeomorphisms, by taking a suitable partial gauge
fixing, to symplectomorphisms preserving a symplectic two-form
constructed from the antisymmetric field of the 5-brane. We will
show there is no local dynamics associated to the antisymmetric
field. This one only  contributes to global degrees of freedom. We
finally use this partial gauge fixing to construct a
regularization for the M5-brane in terms of a multi 1-brane theory
with gauge symmetry $SU(N)\times SU(N)$, $N\rightarrow\infty$. The
analogous construction for the $D=10$ 4-brane yields a
regularization in terms of a multi 0-brane theory, but without any
residual gauge symmetry.

\section{The Algebraic Structure of the M5-Brane Hamiltonian}\label{Hamiltonian}

We start recalling the M5-brane Hamiltonian for the bosonic sector
in the light cone gauge that was obtained in
\cite{DeCastro:2001gp},
\begin{equation}\label{hp}
{\ch}_p=\frac{1}{2}\Pi^M\Pi_M+2g+l^{\mu \nu}l_{\mu\nu}+\Theta_{5i}\Omega^{5i}+\Theta_j\Omega^j
+\Lambda^{\alpha\beta}\Omega_{\alpha\beta},
\end{equation}
where
\begin{equation}\label{l}
l^{\mu\nu}=\frac{1}{2}(P^{\mu\nu}+{\H}^{\mu\nu})
\end{equation}
and
\begin{eqnarray}
{\H}^{\mu\nu}&=& \frac{1}{6} \epsilon^{\mu\nu\gamma\delta\lambda}
H_{\gamma\delta\lambda}¥\\
H_{\gamma\delta\lambda}&=& \p_{\rho}B_{\lambda\sigma}
+\p_{\sigma}B_{\rho\lambda} + \p_{\lambda}B_{\sigma\rho}
\end{eqnarray}
$\Theta_{5i}$, $\Theta_j$, $\Lambda^{\alpha\beta}$ are the
Lagrange multipliers associated to the remaining constraints
\begin{eqnarray}
\Omega^{5i}&=&P^{5i}-\H^{5i}=0\label{fclass1}\\
\Omega^{j}&=&\partial_{\mu}P^{\mu j}=0\label{fclass2},\quad
j=1,2,3,4.
\end{eqnarray}
\begin{equation}\label{fclass3}
\Omega_{[\alpha\beta]}=\partial_{[\beta}\{
\frac{1}{\sqrt{W}}[\Pi_M\partial_{\alpha]}{X}^M+\frac{1}{4}V_{\alpha]}]\}.
\end{equation}
where
\begin{equation}
V_\mu=\epsilon_{\mu\alpha\beta\gamma\delta}
l^{\alpha\beta}l^{\gamma\delta},
\end{equation}
(\ref{fclass1}) and (\ref{fclass2}) are the first class
constraints that generate the gauge symmetry associated to the
antisymmetric field and (\ref{fclass3}) is  the volume preserving
constraint. $\sqrt{W}$ is a scalar density introduced by the LCG fixing
procedure, it may be interpreted as the square root of the determinant of an
intrinsic metric over the spatial world volume. $P^{\mu\nu}$ and 
$\Pi_M$ are the 
conjugate momenta to $B_{\mu\nu}$ and ${X}^M$, respectively.  In our notation
caps Latin letters are transverse light cone gauge indices
$M,N=1,\ldots,9$, Greek ones are spatial world volume indices
ranging from $1$ to $5$, and small Latin letters denote spatial
world volume indices from $1$ to $4$.

The elimination of second class constraints from the formulation in
\cite{Pasti1},\cite{Pasti2} and \cite{JS3} to produce a canonical Hamiltonian with
only first class constraints, was achieved at the price of loosing the
manifest 5 dimensional spatial covariance. In this way , the spatial
world volume splits into $M_{5}= M_{4}\times M_{1}$. We will exploit
this decomposition in our analysis of the Hamiltonian. The
supersymmetric version of this theory was given in  \cite{DeCastro:2002vd}
and \cite{congreso}.

We will see now how a formulation in terms of Nambu-Poisson brackets
arises naturally from the analysis of the volume preserving
diffeomorphisms in more detail.

The algebra of the first class constraints $\Omega_{\alpha\beta}$ on
$M_{5}¥$ is

\begin{equation}
\{\langle\xi_{1}^{\alpha\beta}\Omega_{\alpha\beta}\rangle,\langle\xi_{2}^{\mu\nu}\Omega_{\mu\nu}\rangle
\}= \langle 
(\xi_{1}^{\alpha}\xi_{2}^{\beta}-\xi_{1}^{\beta}\xi_{2}^{\alpha})\Omega_{\alpha\beta}\rangle
\end{equation}

where
\begin{equation}
\xi^{\alpha} = \frac{1}{\sqrt{W}}\p_\beta(\sqrt{W}\xi^{\alpha\beta})
\end{equation}
$\xi^{\alpha\beta}$ are the antisymmetric parameters of the
infinitesimal transformation and
\begin{equation}
\langle \bullet \rangle = \int \bullet \sqrt{W} d^{5}\sigma
\end{equation}

The transformation of scalar fields $X^{M}$ under these infinitesimal
diffeomorphisms  is
\begin{equation}
\delta
X^{M}=\{X^M,\langle\xi^{\alpha\beta}\Omega_{\alpha\beta}\rangle
\}= - \xi^{\alpha}\p_\alpha X^{M}
\end{equation}
and that one for their corresponding conjugate momenta is
\begin{equation}
\delta \Pi_M=
\{\Pi_{M},\langle\xi^{\alpha\beta}\Omega_{\alpha\beta}\rangle \}=
- \sqrt{W}\xi^{\alpha}\p_\alpha \frac{\Pi_{M}}{\sqrt{W}}
\end{equation}
since $\Pi_{M}$ are scalar densities.

We notice that the parameter
$\xi^\alpha=\frac{1}{\sqrt{W}}\p_\beta(\sqrt{W}\xi^{\alpha\beta})$
satisfies
\begin{equation}
{\p}_\alpha (\sqrt{W} \xi^{\alpha})=0
\end{equation}
ensuring that
\begin{equation}
 \delta (\sqrt{W} (\sigma))=0
\end{equation}¥
that is , the diffeomorphisms generated by $\Omega_{\alpha\beta}$
are the volume preserving ones over $M_{5}$. All the above
mentioned properties are in general valid for the volume
preserving diffeomorphisms in any p-dimensional world volume. In
the case $p=2$ an explicit solution for $\xi^{\alpha}$ is
\begin{equation}
 \xi^{\alpha}= \frac{\epsilon^{\alpha \beta}}{\sqrt{W}}\p_{\beta} \xi,
\end{equation}
then the transformation rule for any scalar field $\Phi$ becomes
\begin{equation}
    \delta \Phi = \{ \xi,\Phi\}
 \end{equation}
 where
 \begin{equation}
   \{ A,B\} \equiv \frac{\epsilon^{\alpha \beta}}{\sqrt{W}}\p_{\alpha} A \p_{\beta}
   B
 \end{equation}¥
is a Poisson bracket with symplectic two-form
$\omega_{\alpha\beta} = \sqrt{W}\epsilon_{\alpha\beta}¥$. The
structure constants of the area preserving diffeomorphisms, in the
$p=2$ case, were interpreted as the $N \rightarrow \infty$ limit
of $SU(N)$ in \cite{deWit:1990vb}, leading to a $SU(N)$
regularization of the $D=11$ Supermembrane and its further quantum
mechanical analysis of the canonical Hamiltonian
\cite{deWit:1989ct}. In $p=2$, the area preserving diffeomorphisms
are exactly the same as the symplectomorphisms preserving
$\omega_{\alpha\beta}$. This property of the $p=2$ case was an
essential ingredient in the noncommutative formulation of $D=11$
Supermembranes presented in \cite{Martin:2001zv} and of its
quantum mechanical analysis in \cite{Boulton:2001gz}
\cite{Boulton:2002br}.For $p\geq 2$, the symplectomorphisms are
only a subgroup of the volume-preserving diffeomorphisms.

We may now consider an explicit solution for $p\geq 2$. There always
exist $\xi_{1},\ldots \xi_{p-1}$ such that
\begin{equation}
\xi^{\nu}=\ds\frac{\epsilon^{\nu\alpha_1\ldots\alpha_{p-1}}}{\sqrt{W}}\p_{\alpha_1}\xi_1
\ldots\p_{\alpha_{p-1}}\xi_{p-1},
\end{equation}
consequently, the volume preserving transformation of any scalar field
over the p-brane world volume becomes
\begin{equation}\label{NPbracket}
\delta \Phi
= - \frac{\epsilon^{\mu\alpha_1\ldots\alpha_{p-1}}}{\sqrt{W}}\p_{\mu}\Phi\p_{\alpha_1}\xi_1
\ldots\p_{\alpha_{p-1}}\xi_{p-1}
\end{equation}
which leads to the introduction of the Nambu bracket \cite{Nambu:1973qe} with
p-entries.
\begin{equation}
\delta \Phi= - \{\Phi,\xi_1,\xi_2,\xi_3,\ldots,\xi_{p-1}\}
\end{equation}

It turns out that the commutator of the volume preserving
diffeomorphisms may now be expressed in terms of the Nambu algebraic
structure:
\begin{equation}\label{bracket2}
[{\hd},\delta]\Phi= \tilde{\delta}\Phi
\end{equation}
where the parameters may be expressed as
\begin{equation}\label{bracket1}
\{\hx_i,\xi_1,\xi_2,\xi_3,\cdots,\xi_{p-1}\}=\tilde{\xi}_i.
\end{equation}

It is interesting that the algebraic properties leading to
(\ref{bracket2}) and (\ref{bracket1})¥ arise only from the
generalized Jacobi identity for n-Lie algebras defined by
skew-symmetric n-brackets, $n\leq p$ acting on the ring of
$C^{\infty}$ functions, with no need of an explicit expression for
the Nambu-Poisson bracket. The fundamental identity or generalized
Jacobi identity \cite{Fi}is
\begin{widetext}
\begin{eqnarray}\label{FI}
&&\{\{ f_n, f_1,f_2,f_3, \cdots, f_{n-1}\},\; f_{n+1},
f_{n+2},f_{n+3}\cdots , \;f_{2n-1}\}\cr\cr &&\quad +\; \{ f_n, \{
f_{n+1}, f_1,f_2,f_3, \cdots, f_{n-1}\}, f_{n+2},f_{n+3}, \cdots,
f_{2n-1}\}\cr\cr &&\quad +\;\cdots\; +\; \{ f_n,f_{n+1},f_{n+2},
\cdots f_{2n-2}, \{ f_{2n-1}, f_1,f_2,f_3, \cdots,
f_{n-1}\}\}\cr\cr &&\quad =\;\{\{ f_n, f_{n+1},f_{n+2},\cdots,
f_{2n-1}\}, f_1,f_2,f_3, \cdots, f_{n-1}\}\;.
\end{eqnarray}
\end{widetext}

For any bracket satisfying this identity (\ref{FI}) and the
Leibnitz rule \cite{Takhtajan:1994vr}, there exists an n-vector field
$\bm{\omega}$ such that
\begin{equation}
\{f_1,f_2,\ldots,f_n\}=\bm{\omega}^{\mu_1,\mu_2,\ldots,\mu_n}\p_{\mu_1}f_1\ldots\p_{\mu_n}f_n.
\end{equation}
Any bracket satisfying (\ref{FI}) and the Leibnitz rule is called
a Nambu-Poisson bracket. When $n\geq 3$,  $\bm{\omega}$ is
decomposable at its regular points, that is
\begin{equation}
\bm{\omega}^{\mu_1,\mu_2,\ldots,\mu_n}=\frac{\epsilon^{\mu_1,\ldots,\mu_n\rho_1\ldots,\rho_m}}{w}\p_{\rho_1}g_1\ldots\p_{\rho_m}g_m
,
\end{equation}
this means that any Nambu-Poisson bracket may be expressed at its
regular points as a Nambu bracket \cite{Grabowski1}
\begin{equation}
\{f_1,\ldots,f_n\}_{NP}=\{f_1,\ldots,f_n,g_{1}\ldots g_{p-n}\}
\end{equation}
for some $g_{1}\ldots g_{p-n}$.

Moreover, an n-bracket with $n\geq 3$ entries is Nambu-Poisson if
and only if by fixing an argument one obtains an $(n-1)$
Nambu-Poisson bracket \cite{Grabowski2}, related work may be found in 
\cite{Simon}. This is the kind of
bracket we will find in the Hamiltonian of the M5-brane.

We conclude then that it is possible to endow any p-brane world
volume with a Nambu-Poisson structure. In \cite{hopetesis} it was
proposed such kind of algebraic structure for world volume
p-branes.  The Nambu-Poisson structure over the world volume
coexists with the canonical structure of the field theory.

\section{The Nambu-Poisson structure of the M{5}-Brane Hamiltonian}\label{Nambu}

 In two dimensions, the area preserving
diffeomorphisms are the same as the symplectomorphisms. This
feature has very interesting consequences for the supermembrane
since the intrinsic symplectic structure arising, for example,
from a non-trivial central charge in the supersymmetric algebra
gives then rise to a formulation of the theory, in terms of a
non-commutative geometry \cite{Martin:2001zv}.  In higher
dimensions,  the symplectomorphisms are a subgroup of the full
volume preserving diffeomorphisms. It is then  interesting to
analyze how the geometry of the  M5-brane allows the reduction
from the volume-preserving diffeomorphisms to only
symplectomorphisms . In the supermembrane case it is assumed that
the spatial part of the world-volume is a Riemann surface. Those
are complex manifolds which admit a symplectic structure,
moreover, they are K\"{a}hler manifolds. In our formulation the
spatial five dimensional world-volume $M_5$ has the structure
$M_4\times M_1$. This is required to eliminate second class
constraints. We will assume that  $M_4$ admits a symplectic
structure denoted as $\omega^0$. It is convenient to identify the scalar density $\sqrt{W}$
with the one arising from the symplectic structure over $M_4$.

 We will now show how the interacting terms of the M5-brane Hamiltonian (\ref{hp}) may be expressed
directly in terms of the Nambu-Poisson bracket in five dimensions.
Let us analyze it term by term. We first notice that $g$,
the determinant of the induced metric may be re-expressed in a
straightforward manner as a bracket
\begin{eqnarray}
g&=&\frac{1}{5!}\epsilon^{\nu_1,\ldots,\nu_5}\epsilon^{\mu_1,\ldots,\mu_5}g_{\mu_1\nu_1}\ldots
g_{\mu_5,\nu_5}\cr &=&\frac{1}{5!}\{X^M,X^N,X^P,X^Q,X^R\}^2.
\end{eqnarray}
Let us consider now the third term, it depends on the
antisymmetric field $B_{\mu\nu}$. It is invariant under the action
of the first class constraints (\ref{fclass1}) and
(\ref{fclass2}). To eliminate part of these constraints, we
proceed to make a partial gauge fixing on $B_{\mu\nu}¥$, following
\cite{DeCastro:2001gp} we take
\begin{equation}
B_{5i}=0
\end{equation}
which, together with the constraint (\ref{fclass2}), allow us a
canonical reduction of the Hamiltonian (\ref{hp}). We notice that the
contribution of this partial gauge fixing to the functional
measure is $1$.  We are then left with the constraint
\begin{equation}\label{fclass4}
\partial_j P^{ij}+\partial_5{\H}^{5i}=0 \quad\quad i,j=1,2,3,4,
\end{equation}
which generates the gauge symmetry on the two-form $B$

\begin{equation}
\delta B_{ij}=\partial_i\Lambda_j-\partial_j\Lambda_i.
\end{equation}
$B$ as a two-form over $M_4$ may be  decomposed using the Hodge
decomposition theorem in an exact form plus a co-exact form plus
an harmonic form. With this at hand,  an admissible gauge fixing 
consists to impose the exact form part to be zero. In this gauge, we may express
$P^{ij}$ as
\begin{equation}P^{ij}=\epsilon^{ijkl}(\frac{1}{2}\partial_5
B_{kl}+\widetilde{\omega}_{kl})\end{equation} where
$\widetilde{\omega}$ is a closed two form. We also
obtain
\begin{eqnarray}
l^{ij}&=&\epsilon^{ijkl}(\partial_5
B_{kl}+\widetilde{\omega}_{kl})\cr
l^{5i}&=&\epsilon^{ijkl}\partial_{j}B_{kl},
\end{eqnarray}

The kinetic term in the action
associated to the antisymmetric field becomes then
\begin{equation}
\langle P^{ij}\dot{B}_{ij} \rangle=\langle
\epsilon^{ijkl}\dot{B}_{ij}\widetilde{\omega}_{kl} \rangle,
\end{equation}
where the co-exact part of $B_{ij}$ is conjugate to the exact part
of $\widetilde{\omega}_{kl}$, while the harmonic parts are
conjugate to each other.

Noticing that $l^{5i}$ is divergenceless, it may be rewritten as
\begin{equation}\label{decom2}
l^{5i}=\epsilon^{5ijkl}\partial_j\phi_{[a}\partial_k\phi_b\partial_l\phi_{c]},
\quad\quad a,b,c=1,2,3.
\end{equation}

This decomposition in terms of scalars is always valid for any
four dimensional divergenceless smooth vectorial density. The
co-exact part of $B_{kl}$ may then be expressed in a unique way,
modulo the symmetry generated by the group of rigid
transformations that leaves invariant (\ref{decom2}), in terms of
the triplet $\phi_{a}$.

Now we decompose the tensor density $l^{ij}$ into
\begin{equation}\label{decom1}
l^{ij}=\epsilon^{ji\alpha\beta\gamma}
\partial_\alpha\phi_{[a}\partial_\beta\phi_b\partial_\gamma\phi_{c]}+\epsilon^{jikl}\omega_{kl}
\end{equation}
where  $\omega$ is a closed two form. Comparing
$\widetilde{\omega}$ with $\omega$, we notice that part of
$\widetilde{\omega}$ has been absorbed into the first term of the
right hand side of (\ref{decom1}). This decomposition exists and
it is unique.

It is now possible, as a consequence of the Darboux's theorem , to express
$\omega_{kl}$ in terms of the  two-form $\omega^{0}$ over
$M_{4}$
\begin{equation}\label{decom3}
\omega_{jl}=\partial_j\Psi^k\partial_l\Psi^m \omega^0_{km}
\end{equation}
with $k,m=1,2,3,4$. When $\omega$ is non degenerate, there exists
an atlas $\{ U\}$ such that, on each open set $U$, $\Psi^k$
describe the diffeomorphisms which reduces $\omega_{jl}$ to the two-form $\omega^0$. When $\omega$ is
degenerate the decomposition (\ref{decom3}) is still valid. In
that case pairs of  $\Psi^k$ may be zero.

Although there are four fields $\Psi^k$, they
represent only three physical degrees of freedom, since on an open
set where $\omega$ has constant rank , one may reduce locally $\omega$ to $\omega^0$ by a change of variables and still have one gauge
symmetry left, the diffeomorphisms which preserve $\omega^0$. The
infinitesimal group parameter of those diffeomorphisms  can be
expressed as
\begin{equation}
\xi^i= 4\omega^0_{kl}\frac{\epsilon^{ijkl}}{\sqrt{W}}\partial_j\xi.
\end{equation}

Using (\ref{decom2}), (\ref{decom1}) and (\ref{decom3}) the 
interacting term in the Hamiltonian  involving the antisymmetric field and its
conjugate momentum may be expressed in terms of the Nambu-Poisson
brackets
\begin{eqnarray}
g_{\alpha\mu}g_{\beta\nu}l^{\alpha\beta}
l^{\mu\nu}&=&(l^{\mu\nu}\partial_\mu X^M\partial_\nu X^N)^2\\
&=&(\{\phi_a,\phi_b,\phi_c,X^M,X^N\}\epsilon^{abc}\\ &
+&\{\Psi^k,\Psi^l,X^M,X^N\} \omega^0_{kl})^2.
\end{eqnarray}

Moreover, in the case of a non degenerate $\omega$, the second
Nambu-Poisson bracket may be re-expressed, on any open set of a Darboux atlas,
in terms of a Poisson bracket constructed with the symplectic two-form
$\omega^0_{ij}$ by fixing the volume preserving
diffeomorphisms:
\begin{equation}
(l^{\mu\nu}\p _{\mu} X^{M}\p _{\nu}
X^{N})^{2}=(\{\phi_a,\phi_b,\phi_c,X^M,X^N\}\epsilon^{abc}
+\{X^M,X^N\})^2
\end{equation}
where
\begin{equation}
\{X^M,X^N\}=\frac{4\epsilon^{ijkl} \omega^{0}_{kl}}{\sqrt{W}¥}\partial_i X^M\partial_j X^N
\end{equation}

In this case, the Hamiltonian is still invariant under the
symplectomorphisms which preserves $\omega^0$. The interacting terms 
of the canonical Lagrangian may then be expressed in terms of
the Nambu-Poisson bracket constructed with the antisymmetric
tensor ${\epsilon^{\mu\nu\alpha\beta\gamma}}/{\sqrt{W}}$ and the
Poisson bracket constructed with the symplectic two-form
$\omega^0$.

\section{M5-brane and D4-brane regularization in terms of D0 and D1-branes}\label{matricial}

 We will discuss in this section a  partial gauge fixing of the M5
brane which yields a formulation of the theory invariant under
symplectomorphisms  only, allowing  a regularization
with a similar approach to the one of the $D=11$ supermembrane
\cite{deWit:1989ct}.In here, we will assume $\omega$ to be non-degenerate.
After fixing $\omega$ to $\omega^0$ we may resolve the
volume-preserving constraint for $\phi_{a}$ $a=1,2,3$. We are then
left still with one constraint,
\begin{equation}
\epsilon^{ijkl} \omega^0_{kl}\p_i ( \frac{\Pi_M\p_jX^M}{\sqrt{W}}) =0.
\end{equation}
The left hand member generates the symplectomorphisms preserving
$\omega^0$. The five dimensional volume preserving diffeomorphisms have been
reduced by the gauge fixing procedure to only that generator. We are then left with a
formulation in terms of $X^M$ and its conjugate momenta $\Pi_M$,
invariant under symplectomorphisms. The antisymmetric field
$B_{\mu\nu}$ and its conjugate momenta $P^{\mu\nu}$ have been
reduced to $\omega^0$, there is no local dynamics related to them. All
the dynamics may be expressed in terms of $(X^M,\Pi_M)$. We may
then perform the explicit $4+1$ decomposition  on the spatial
sector of the world-volume. We obtain
\begin{eqnarray}
\{X^M,X^N,X^P,X^Q,X^R\}&=&\p_5X^M\p_5X^N G_{MN}(X)\cr
&=&\p_5X^{[M}\{X^N,X^P,X^{Q},X^{R]}\}.
\end{eqnarray}
We may also rewrite the Nambu-Poisson bracket in terms of the
Poisson bracket as
\begin{equation}
\{X^N,X^P,X^Q,X^{R}\}= 8 \{X^{[N},X^P\}\{X^{Q]},X^{R}\}
\end{equation}
where the brackets on indices denote cyclic permutation.

The contribution of the other terms in the Hamiltonian may be
expressed in a similar way. The scalar fields $\phi_{a}$ may be 
eliminated from the volume preserving constraints. This elimination is 
algebraically possible since they appear linearly in them. We will show the
complete analysis for the D4-brane in $10$ dimensions. The
Hamiltonian may be obtained from the one for the M5-brane by
considering the potential gauge fixing $X^{11}=\sigma^5$, together
with the assumption that all other fields are independent of
$\sigma^5$. $\pi_5$ is eliminated from one of the constraints.

The resulting Hamiltonian acquires then, as a result of the
$\Pi_5$ elimination, a quadratic term in $P^{ij}$, $i,j=1,2,3,4$.
The bosonic sector of the Hamiltonian may then be expressed as
\begin{equation}\label{nambuhamil}
{\ch}=\frac{1}{2}\Pi^2+g+(\H^i\p_iX^M)^2+\{X^M,X^N\}^2+(\epsilon_{ijkl}P^{ij}P^{kl})^2.
\end{equation}
The last term on the right hand side is, in this case, the
determinant of the Riemannian metric constructed from the
symplectic structure and the integrable almost complex structure
of the  manifold $M_4$. It only contributes to the global degrees
of freedom. In the case of the D4-brane in 10 dimensions, the
residual gauge freedom after fixing the LCG is again the volume
preserving diffeomorphisms. However, in distinction, to what
occurs in the M5-brane, we cannot apply directly the Darboux
argument for a closed, non-degenerate two-form, since the 4
dimensional diffeomorphisms are restricted by the volume
preserving condition. However, one can show that the Darboux map
from $\omega$ to $\omega^0$ exists provided the scalar field
\begin{equation}\label{scalar}
\frac{\epsilon^{ijkl}\omega^0_{ij}\omega^0_{kl}}{\sqrt{W}}
\end{equation}
is not constant. There are then no residual diffeomorphisms
preserving $\omega^0$ and the volume element $\sqrt{W}$. We thus
end up with a gauge fixed formulation in terms of $X^M$ and
$\Pi_M$. The second term of (\ref{nambuhamil}) is again expressed
in terms of the Nambu-Poisson structure
\begin{equation}
g=\{X^M,X^N,X^P,X^Q\}^2=(8 \{X^{M},X^{[N}\}\{X^P,X^{Q]}\})^2
\end{equation}
while the third term requires the solution of the volume
preserving  constraints. We obtain
\begin{equation}
\H^i=  \frac{4\epsilon^{ijkl} \omega^{0}_{kl}}{\sqrt{W}}(\sqrt{W}\p_j\theta-\Pi_M\p_jX^M)
\end{equation}
where $\theta$ is a scalar field which becomes determined from the
divergenceless property of $\H^i$
\begin{equation}
\{\sqrt{W},\theta\}=\{\Pi_M,X^M\}
\end{equation}
the left hand member is the derivative of $\theta$ in the
direction of the gradient of the scalar field(\ref{scalar}).

The third term of (\ref{nambuhamil}) may thus be written as
\begin{equation}
(\H^i\p_i X^M)^2=(\sqrt{W}\{\theta,X^M\}-\Pi_N\{X^N,X^M\})^2
\end{equation}

All the interacting terms of the Hamiltonian (\ref{nambuhamil})
can then be expressed in terms of the Poisson bracket. 

Finally, in order to obtain a regularization of the above Hamiltonian, we express $X^M$ and $\Pi_M$ in terms of a complete
orthonormal basis over $M_4$, $\{Y_a\}$, in the Hilbert space of
$L^2$ functions. Since
\begin{equation}
\frac{4\epsilon^{ijkl} \omega^{0}_{kl}}{\sqrt{W}¥}\p_iY_a\p_jY_b
\end{equation}
is a scalar function over $M_4$, we may reexpress it in terms of
the basis, and obtain
\begin{equation}
\{Y_a,Y_b\}=\frac{4\epsilon^{ijkl} \omega^{0}_{kl}}{\sqrt{W}¥}\p_iY_a\p_jY_b=f_{abc}Y_c
\end{equation}
where $f_{abc}$ is  given by
\begin{equation}
\int_{M_4}\frac{4\epsilon^{ijkl} \omega^{0}_{kl}}{\sqrt{W}¥}\p_iY_a\p_jY_b Y_c=f_{abc}
\end{equation}
and is totally antisymmetric on $a,b,c$

We then have
\begin{equation}
X^M=X^M_a(\sigma^5,\tau)Y_{b}(\sigma^1,\sigma^2,\sigma^3,\sigma^4),
\end{equation}
consequently,
\begin{equation}
\{X^M,X^N\}= X^M_a X^N_b \{Y_a,Y_b\} = X^M_a X^N_b f_{abc}Y_c.
\end{equation}
Furthermore, we may introduce
\begin{equation}
Y_aY_b=C_{abd}Y_d
\end{equation}
Which is again valid since $Y_a Y_b$ is also a scalar function
over $M_4$. We get
\begin{equation}
\int_{M_4}Y_aY_bY_d=C_{abd}
\end{equation}
which becomes totally symmetric in $a,b,d$.

All the interacting terms in the Hamiltonian may be rewritten using 
$X^M_a$, their conjugate momenta  and the structure constants $f_{abc}$ 
and $C_{abd}$. As an example, the second term in the Hamiltonian 
(\ref{hp}) is given by
\begin{equation}
\int d\tau d\sigma^5\int_{M_4}d\sigma^4 g=\int(\p_5X^{[M}_aX^N_b
X^P_c X^Q_d X^{R]}_e)f_{bcm}f_{den}C_{mnl}C_{alf})^2 d\tau
d\sigma^5
\end{equation}
which now takes the form of the potential of a condensate
D1-branes over a non trivial background. The geometry of the
background depends on the symplectic structure of the $M_4$
manifold. We may then regularize the
Hamiltonian by taking a truncation in the range of the index $a$.
The truncated Hamiltonian for the M5 brane may then be interpreted as a 
$SU(N)\times SU(N)$ gauge theory. While the D4-brane Hamiltonian 
corresponds to a gauge fixed $SU(N)\times SU(N)$ theory. The potential contains powers of
$X^M_a$ up to $8$, moreover, the quadratic term in the momenta is
contracted with a nontrivial background metric. Nevertheless, the
corresponding quantum mechanical problem seems reasonable to be
analyzed. This regularized version of the D4-brane in terms of
D0-branes as well as the one for the M5-brane in terms of
D1-branes are both formulated in terms of $X^M$ and $\Pi_M$ as in
the case of the supermembrane. In the latter case and that one for
the M5-brane , there is a gauge freedom generated by
symplectomorphisms. The antisymmetric field on the M5 and
D4-branes only contributes with global degrees of freedom, not
with local ones.

\section{Conclusions}

We introduced a Nambu-Poisson structure in the description of the
$D=11$ M5-brane and in the $D=10$ D4-brane. By partial gauge fixing the volume preserving
diffeomorphisms, we showed that the symplectic sector of the  M5-brane Hamiltonian may be expressed as
a $D=11$ p-brane theory invariant under symplectomorphisms preserving a 
symplectic two-form
constructed from the antisymmetric field of the M5-brane. All the
degrees of freedom of the antisymmetric field reduce to global ones. We then
propose a regularization of the M5-brane in terms of a multi
D1-brane theory invariant under $SU(N)\times SU(N)$ in the limit
when $N\rightarrow\infty$. We also propose a regularization for
the $D=10$ D4-brane in terms of a multi D0-brane theory.

\begin{acknowledgments}
We would like to thank  Drs. Sim\'on Codriansky, Jaime Camacaro,
and Jens Hoppe for very useful discussions. This research is
supported by Instituto Venezolano de Investigaciones Cient\'{\i}ficas
(IVIC) and Decanato de Investigaciones de la Universidad Sim\'on
Bol\'{\i}var. M.P. Garcia del Moral wishes to thank IFT (UAM-CSIC) for financial support and
kind hospitality while part of this work was done.
\end{acknowledgments}

%\newpage %Just because of unusual number of tables stacked at end
\bibliography{tesbib}% Produces the bibliography via BibTeX.

%--------------------------------------------------------------------

\end{document}